\documentclass[sigconf,screen,natbib]{acmart}
\usepackage{booktabs}
\usepackage{listings}
\usepackage{multirow}
\usepackage{tcolorbox}
\usepackage{tabularx}

\AtBeginDocument{%
\providecommand\BibTeX{{%
\normalfont B\kern-0.5em{\scshape i\kern-0.25em b}\kern-0.8em\TeX}}}

\copyrightyear{2020}
\acmYear{2020}
\setcopyright{acmlicensed}
\acmConference[MSR '20]{17th International Conference on Mining Software Repositories}{October 5--6, 2020}{Seoul, Republic of Korea}
\acmBooktitle{17th International Conference on Mining Software Repositories (MSR '20), October 5--6, 2020, Seoul, Republic of Korea}
\acmPrice{15.00}
\acmDOI{10.1145/3379597.3387486}
\acmISBN{978-1-4503-7517-7/20/05}

\makeatletter

\usepackage{amssymb}
\newcommand{\ygg@basicalert}[2]{\fbox{\bfseries\sffamily\scriptsize#1}{\sf\small$\blacktriangleright$\textit{#2}$\blacktriangleleft$}}
\newcommand{\YANN}[1]{\ygg@basicalert{YANN}{#1}}

\newcommand{\projectpage}{\url{https://github.com/game-dev-database/postmortem-problems}}

\setcitestyle{numbers,sort&compress}

\begin{document}

\title{Dataset of Video Game Development Problems}
\titlenote{The dataset is available in the following page: \projectpage. }

\author{Cristiano Politowski}
\email{c_polito@encs.concordia.ca}
\orcid{0000-0002-0206-1056}
\affiliation{%
  \institution{Concordia University}
  \city{Montreal}
  \state{Quebec}
  \country{Canada}
}

\author{Fabio Petrillo}
\email{fabio@petrillo.com}
\orcid{0000-0002-8355-1494}
\affiliation{%
  \institution{Université du Québec à Chicoutimi}
  \city{Chicoutimi}
  \state{Quebec}
  \country{Canada}
}

\author{Gabriel Cavalheiro Ullmann}
\email{gabriel.cavalheiro@sou.unijui.edu.br}
\affiliation{%
  \institution{UNIJUI}
  \city{Santa Rosa}
  \state{RS}
  \country{Brazil}
}

\author{Josias de Andrade Werly}
\email{alu201460044@uniriter.edu.com}
\affiliation{%
  \institution{UniRitter}
  \city{Porto Alegre}
  \state{RS}
  \country{Brazil}
}

\author{Yann-Ga\"el Gu\'{e}h\'{e}neuc}
\email{yann-gael.gueheneuc@concordia.ca}
\orcid{0000-0002-4361-2563}
\affiliation{%
  \institution{Concordia University}
  \city{Montreal}
  \state{Quebec}
  \country{Canada}
}


\begin{abstract}
Different from traditional software development, there is little information about the software-engineering process and techniques in video-game development. One popular way to share knowledge among the video-game developers' community is the publishing of postmortems, which are documents summarizing what happened during the video-game development project. However, these documents are written without formal structure and often providing disparate information. Through this paper, we provide developers and researchers with grounded dataset describing software-engineering problems in video-game development extracted from postmortems. We created the dataset using an iterative method through which we manually coded more than 200 postmortems spanning 20 years (1998 to 2018) and extracted 1,035 problems related to software engineering while maintaining traceability links to the postmortems. We grouped the problems in 20 different types. This dataset is useful to understand the problems faced by developers during video-game development, providing researchers and practitioners a starting point to study video-game development in the context of software engineering. 
\end{abstract}

\maketitle

\section{Introduction}

Video games are a profitable industry \cite{Newzoo19} for big companies like EA\footnote{Eletronic Arts report from October 2019 shows \$3.8 bilions of revenue (\url{http://bit.ly/2LA7us1}).} and for \emph{indie} developers\footnote{Generally, indie developers are small, self-funded companies that create games on a small scale.} brave enough to face a competitive endeavor.
Successful games are plenty, although they usually come at a price. Indeed, video-game development is known for its management problems \cite{Petrillo2009, Washburn2016}, which translate into large numbers of issues in the final products. For example, 80\% of the top 50 games on Steam\footnote{Steam is a video-game, digital, distribution service platform available at \url{https://store.steampowered.com/}.} need critical updates \cite{Lin2017:updates}. 

Video games form also a competitive market in which knowledge is the main weapon against competitors. The lack of information about the process, the techniques, the game engine used for games are but a few examples of how hard it is to understand video-game development. This lack of information also prevents new developers (and even veterans) to avoid common problems by learning from previous mistakes.

One source of information about video-game development that is public are postmortems. Video-game postmortems are documents that summarize the developers' experiences with the game development, written often right after the game is launched \cite{Washburn2016}. These documents are usually written by managers or senior developers \cite{Callele2005}. They often include five sections either about ``what went right'' and ``what went wrong'' during the game development.

\begin{itemize}
\item \emph{``What went right''} discusses the best practices adopted by developers, solutions, improvements, and project-management decisions that helped the project.
\item \emph{``What went wrong''} discusses difficulties, pitfalls, and mistakes experienced by the development team in the project, both technical and managerial.
\end{itemize}

Thus, postmortems offer an ``open and honest window into the development of games'' \cite{Washburn2016}. However, these documents are written without formal structure and often providing disparate information, in particular with respect to the software-engineering of the games. Consequently, complete, trustful information about game development is hard to find, which limits the number of studies about games from the point of view of software engineering. Yet, we believe that most of software-engineering problems could be mitigated if developers had information beforehand. 

Through this paper, we propose a curated dataset describing software-engineering problems in video-game development. We compile a large set of information about problems in game development related to software engineering. We analyze 200 postmortems from the Gamasutra Website\footnote{Gamasutra is the main and most complete source of video-game postmortems: \url{http://www.gamasutra.com}. It is a descendant of the Game Developers Magazine that ended in 2013.} and collected 1,035 problems. We categorized these problems into 20 different problems types and provide a structure to store this information.
The dataset is available in the following page: \projectpage.

The paper is structured as follows. Section~\ref{sec:related-works} discusses papers that used video-game postmortems as a knowledge base. Section~\ref{sec:method} describes how the data was gathered. Section~\ref{sec:metadata} presents how the metadata is organized. Section~\ref{sec:results} shows the results of the database. Section~\ref{sec:conclusion} concludes the paper with future work.

\section{ Game Development Problems}
\label{sec:related-works}

\citet{Callele2005} analyzed 50 postmortems from the Game Developer Magazine and investigated how requirements engineering is applied in game development. They grouped ``What went right'' and ``What went wrong'' in the five categories: (1) \emph{pre-production}, problems outside of the traditional software development process; (2) \emph{internal}, those related to project management and personnel; (3) \emph{external}, those outside of the development team's control; (4) \emph{technology}, those related to the creation or adoption of new technologies; and, (5) \emph{schedule}, those related to time estimates and overruns. They concluded that project management is the greatest contributor to the success or failure in video-game development.

\citet{Petrillo2009} analyzed 20 postmortems from the Gamasutra Website, searching for the most common problems, which they compared with traditional software problems. They concluded that (1) video-game development suffer mostly from management problems instead of technical problems; (2) the problems found in video-game development are also found in traditional software development; and, (3) the most common problems are related to \emph{scope}, \emph{feature creep}, and \emph{cutting features}. 

\citet{Kanode2009} used postmortems to discuss the challenges of traditional software engineering in video-game development. They reported differences between game development and traditional development and concluded that video-game development must adopt and adapt software-engineering practices.

\citet{Lewis2011} used two previous papers \cite{Blow2004, Tschang2005} to identify problems in games and whether/why these could be of interest to the software-engineering research community. They highlighted some areas to explore further and differences between games and traditional software.

\citet{Washburn2016} analyzed 155 postmortems, identified some characteristics and pitfalls, and suggested good practices. They reported the following main problems: \emph{obstacles} (37\%), \emph{schedule} (25\%), \emph{development process} (24\%), and \emph{game design} (22\%). 

\citet{Politowski2016} extracted development processes from 20 postmortems. They concluded that the majority of the game industry uses agile instead of the waterfall method. In their next work, \citet{Politowski2018} used 55 postmortems to create a recommendation system for new video-game projects using previous mistakes and good practices.

From the gray literature, \citet{Shirinian2011} wrote an article analyzing 24 postmortems from 2008 to 2010 and created a short list of definition from ``what went right'' and ``what went wrong'': 
\emph{Design} (covers all situations and decisions that were made that are clearly external to the direct team and development process);
\emph{Art} (Relating to art decisions);
\emph{Production/Process} (This relates to scheduling, work prioritization, etc);
\emph{Programming} (technical issues); and
\emph{Testing} (all traditional QA functions)

These previous papers used postmortems to analyze video-game development, in particular \citet{Petrillo2009} and \citet{Washburn2016}. We extend these papers with more methodical approach and a study of a larger number of postmortems. We study postmortems from the point of view of software engineering using an iterative approach, which we summarise in the following section.

\section{Method of Analysis}
\label{sec:method}

\begin{figure}[ht]
\centering
\includegraphics[width=1\linewidth]{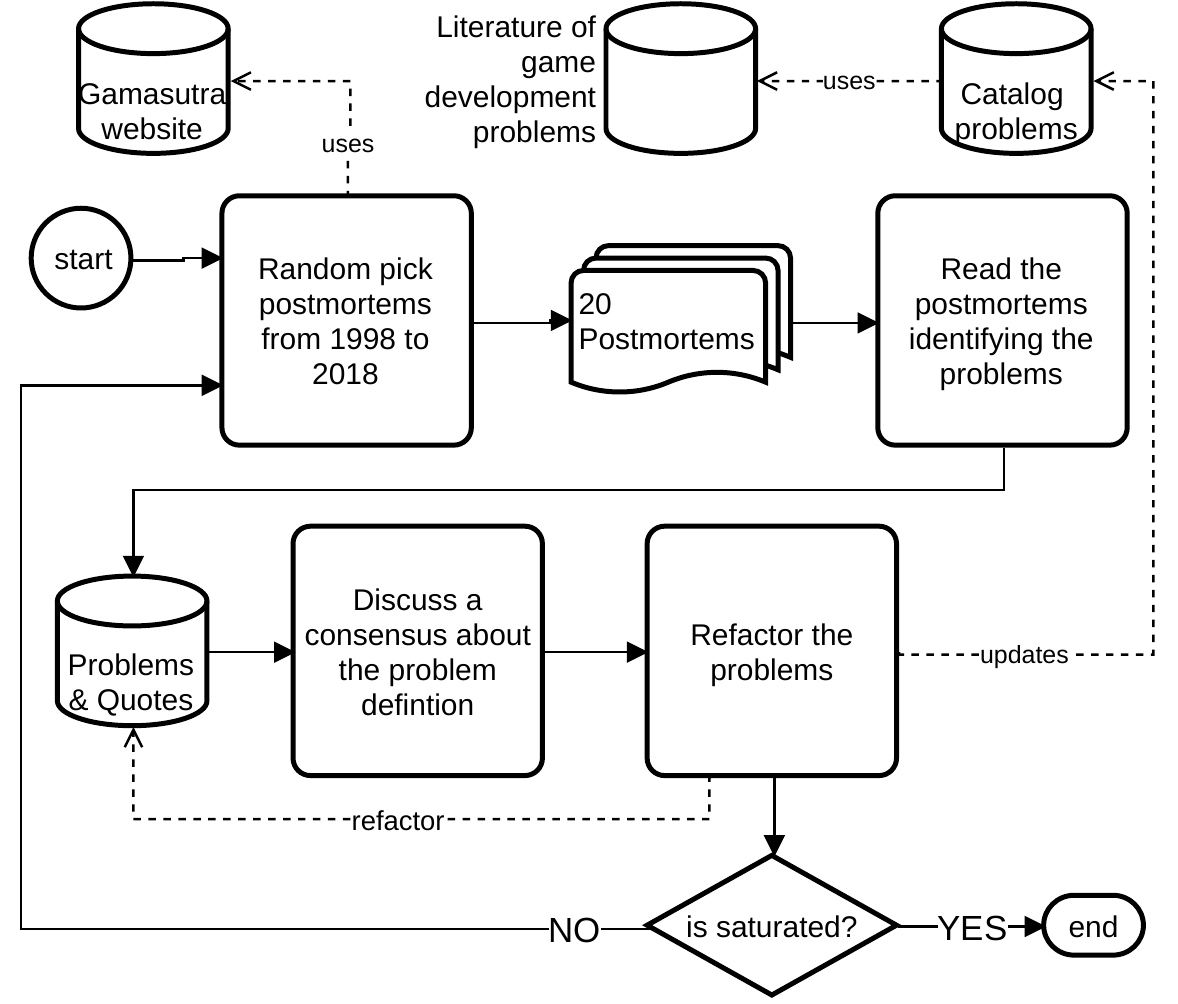}
\caption{Steps performed to analyse the data.}
\label{fig:method}
\end{figure}

We designed the process of analysis to be iterative where the data keep in constant evolution as we add and refactor the data each new iteration. \autoref{fig:method} shows the steps performed to analyse the data from the postmortems.

We start by randomly picking 20 postmortems (for each author) from the Gamasutra Website between the years 1998 to 2018. Next, each author reads the postmortems, focusing on the ``What went wrong'' section and using a coding technique from Grounded Theory \cite{Stol2016}. Each author identifies all the problems reported by the postmortems, extracting quotes and grouping similar problems, based on the previous literature definitions \cite{Petrillo2009,Washburn2016}. Previous definitions were used to create a \emph{catalog of problems}, which contains \emph{problem types}, \emph{problem groups}, and \emph{descriptions}. We add each new problem type absent from the catalog with its description.

In the next step, we discuss the findings, review some doubts, and analyse the correctness of the catalog. Any change at this point, on the catalogue or on the database of problems, results in a update in both documents, which sometimes lead us to re-read the postmortems. We reiterate this process until it reaches saturation: until no more new type of problems appears (stop updating the the catalog of problems).

To keep the distribution of the postmortems read by year balanced, eventually we chose specific postmortems to analyse instead of randomly picking them. Also, not all postmortems are relevant for this study as they might not have useful information regarding the development process. When that happens, we discard the postmortem.

\section{Metadata of the Dataset}
\label{sec:metadata}

The \emph{catalog of problems} is a document listing the possible problems we could find while reading the postmortems.
We started building it using problems' types we gathered from the literature and we update it during during the postmortems analysis.

The catalogue document is updated for every new problem type we end up discovering. The authors review each new problem type and reach to a conclusion.
The final version of the document contains 20 different types of 1,035 problems divided in three groups.

\autoref{tab:catalogue} shows the final version of the catalog where a \emph{Problem Type} is a index to define the problem which has a short \emph{Description} and belongs to one \emph{Problem Group}.

\begin{table}[ht]
\footnotesize
\caption{Catalogue with the video game development problems identified from the postmortem analysis. }
\label{tab:catalogue}
\begin{tabularx}{\linewidth}{@{}>{\hsize=.2\hsize}X>{\hsize=.2\hsize}X>{\hsize=.6\hsize}X@{}}
\toprule
Group & Type & Description of the problem \\ 

\midrule
\multirow{15}{*}{Production} & Design & Any problem regarding the design of the game, like balancing the gameplay. Not a technical detail. \\ \addlinespace
 & Documentation & Not planning the game beforehand, not documenting the code, artifacts or game plan. \\ \addlinespace
 & Tools & Any problem with tools like engines, APIs, development kits, third-party software, etc. \\
\addlinespace & Technical & Problems with the team code/assets infra-structure. \\
\addlinespace & Testing & Any problem regarding the testing. \\
\addlinespace & Bugs & When there are too many bugs in the game/engine, any failure in the game design or technical issues. \\
\addlinespace & Prototyping & Lack of or no prototyping phase nor validation of the gameplay/feature. \\

\midrule 
\multirow{19}{*}{Management} & Unrealistic Scope & Planning too many features that end up impossible to implement it in a reasonable time. \\
\addlinespace & Feature Creep & Adding unplanned new features to the game during its implementation. \\
\addlinespace & Cutting Features & Cutting features previously planned because of other factor like short deadlines. \\
\addlinespace & Delays & Problems regarding any delay in the production. \\
\addlinespace & Crunch Time & When developers continuously spent extra hours working in the project. \\
\addlinespace & Communication & Problems regarding communication with any stakeholder. \\
\addlinespace & Team & Problems in setting up the team, loss of professionals during the development or outsourcing. \\
\addlinespace & Over Budget & Project cost more money than expected. \\
\addlinespace & Multiple Projects & When there is more than one project being developed at the same time. \\
\addlinespace & Planning & Problems involving too much time planing/scheduling or the lack of it. \\
\addlinespace & Security & Problems regarding leaked assets. \\

\midrule
\multirow{3}{*}{Business} & Marketing & Problems regarding marketing/advertising \\
\addlinespace & Monetization & Problems with the process used to generate revenue from a video game product. \\ \bottomrule
\end{tabularx}
\end{table}

\begin{figure}[ht]
\centering
\includegraphics[width=1\linewidth]{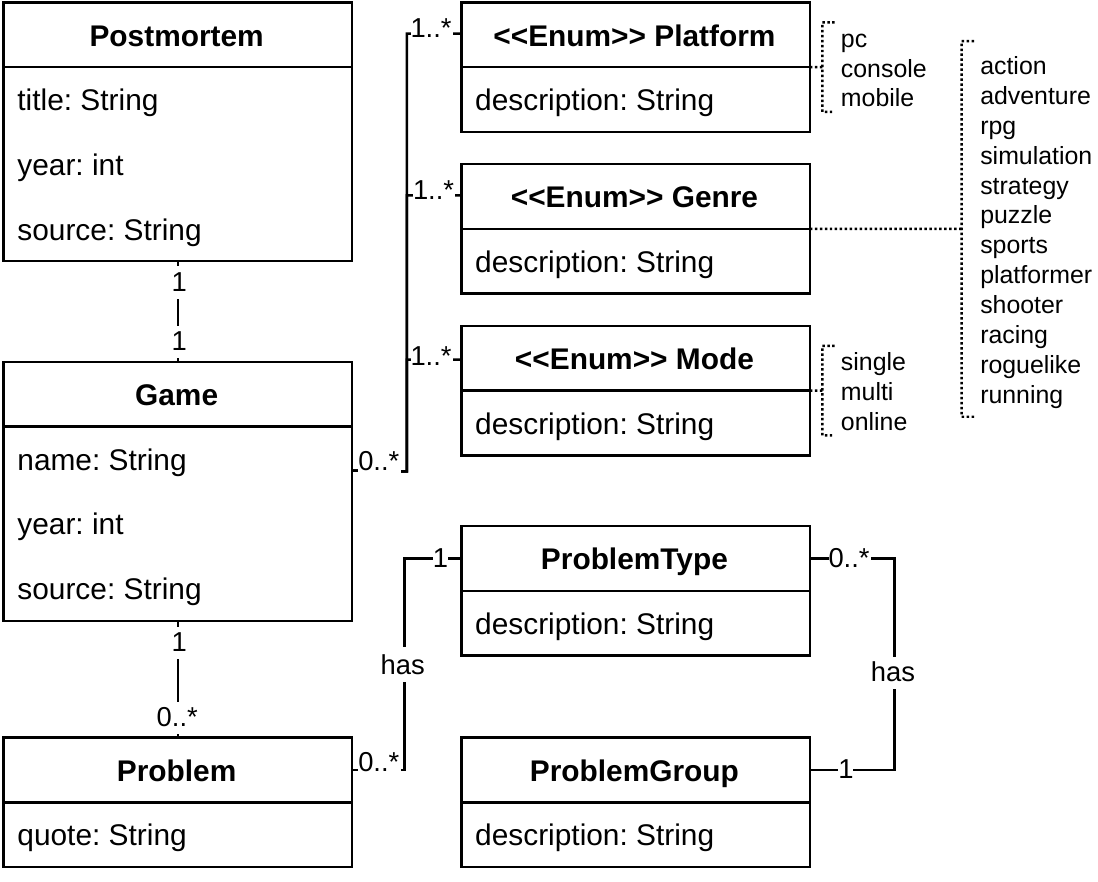}
\caption{Class diagram with the structure of each entry on the dataset.}
\label{fig:structure}
\end{figure}

To store the problems gathered from postmortems we defined a data model which \autoref{fig:structure} shows in UML class diagram. 
Each \textit{Postmortem} has a \textit{Game} which has a collection of problems. Each \textit{Problem} has a \textit{Type} and \textit{Group}. Also, the Game must have: a Platform [1-3] (PC, Console, Mobile), a Genre [1-12] (Action, Adventure, RPG, Simulation, Strategy, Puzzle, Sports, Platformer, Shooter, Racing, Roguelike, Running), and a Mode [1-3] (Single-player, Multi-player, Online).

An example of the dataset structure is shown in \autoref{tab:dataset-structure}. In this case, the game ``Baldur's Gate II'' from 2001 was analyzed and the entry shows a problem regarding ``Testing'', which belongs to the ``Production'' group. The quote relates the problem to a non-modified version taken from the postmortem.

We began storing the data using JSON format
following the structure described in \autoref{fig:structure} and validated using JSON Schema.
Later we migrated the data to a table format (.CSV) where each entry (line) refers to a problem.
This makes the data easy to be analyzed with scripts, specially in \textit{R} language.

\begin{table}[ht]
\footnotesize
\caption{Example of the dataset structure with one entry.}
\label{tab:dataset-structure}
\begin{tabularx}{\linewidth}{@{}rX@{}}
\toprule
Column & Value \\
\midrule
Title & Baldurs Gate II -- The Anatomy of a Sequel \\
Year & 2001 \\
Source & \url{http://bit.ly/2IDsVa0} \\
Name & Baldur's Gate II \\
Platform & PC \\
Genre & RPG Strategy \\
Mode & Multi Single \\
Group & Production \\
Type & Testing \\
Quote & 
(...) We put a number of white-boards in the halls of the testing and design area and listed all of the quests on the boards. We then put an X next to each quest. We broke the designers and QA teams into paired subgroups - each pair (one tester and one designer) had the responsibility of thoroughly checking and fixing each quest. After they were certain the quest was bulletproof, its X was removed. It took about 2 weeks to clear the board (on the first pass). \\
\bottomrule
\end{tabularx}
\end{table}

\section{Dataset analysis}
\label{sec:results}

\begin{figure}[ht]
\centering
\includegraphics[width=1\linewidth]{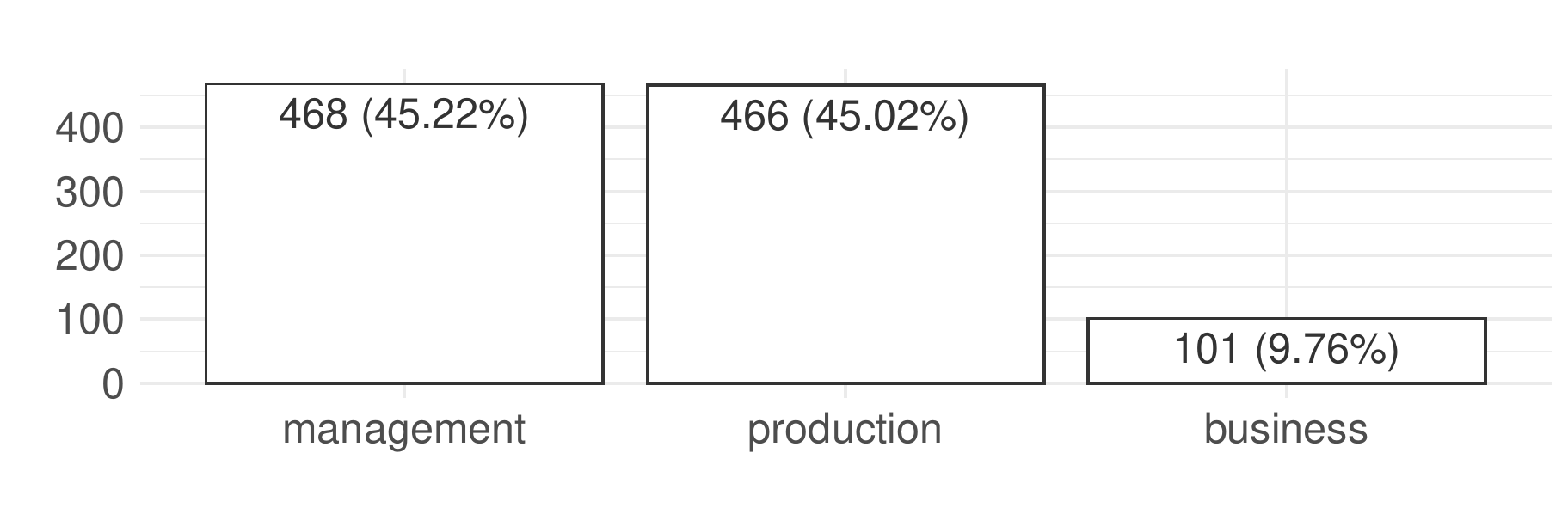}
\caption{Number of problems related to each Group.}
\label{fig:problem-group}
\end{figure}

\begin{figure}[ht]
\centering
\includegraphics[width=1\linewidth]{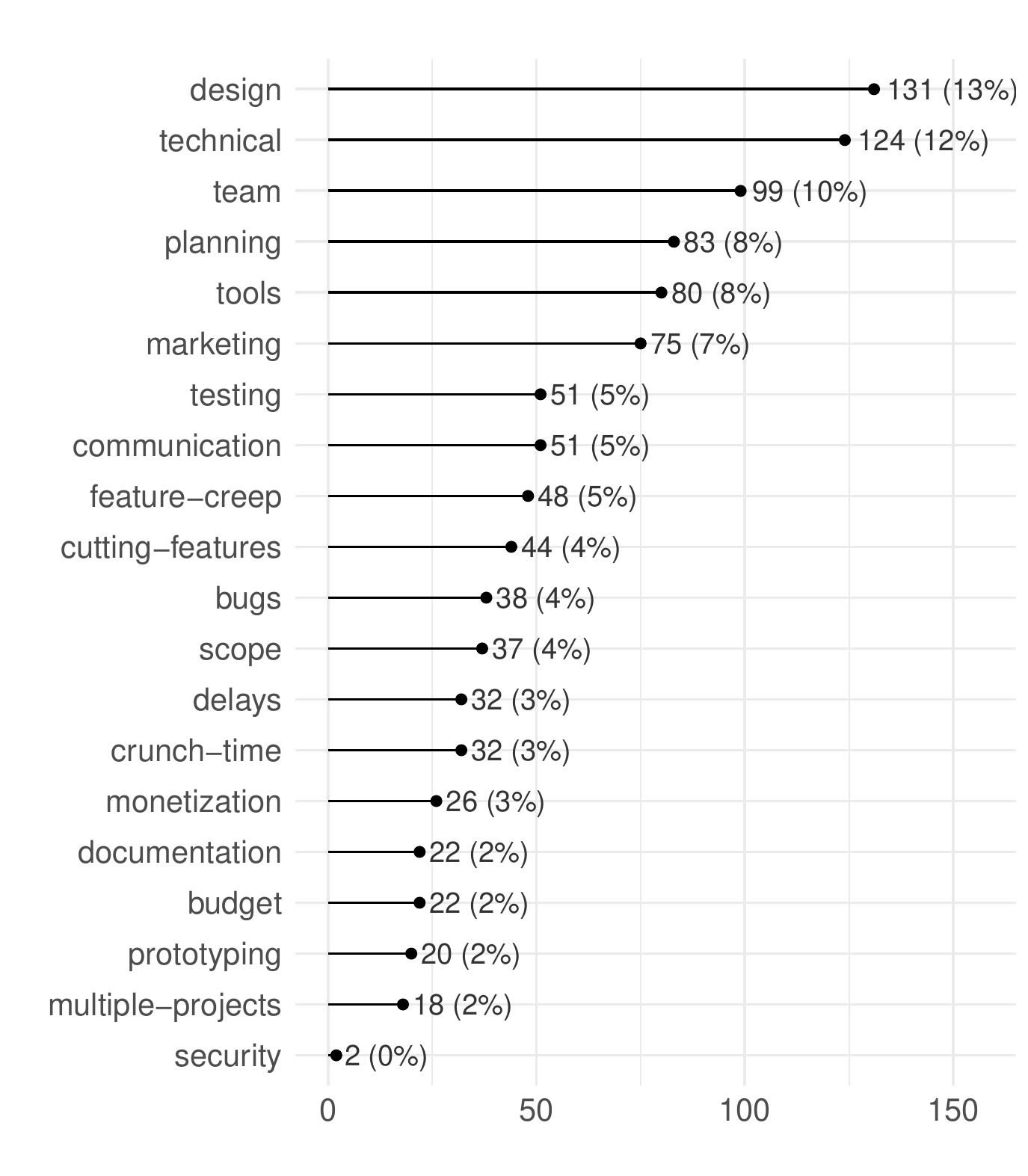}
\caption{Number of problems related to each Type.}
\label{fig:problem-type}
\end{figure}

The dataset contains 200 video-game projects ranging from 1998 to 2018. There are, in total, 1,035 problems with its respective quotes. The median of problems by title (video game) is 5 and the median of problems by year is 48. 
\autoref{fig:problem-group} shows the problems by group: management and production problems are the most common with around 45\% each while business issues sum up to 10\%.
\autoref{fig:problem-type} shows the distribution of the problems related to each type. In this case, Design, Technical, and Team problems are the most frequent with 35\% overall.

Regarding the platforms, the projects are mainly for PC, with 787 problems, followed by 475 Console problems, and 254 Mobile problems. 90 problems are related to multi-platform projects, that is, PC, Console, and Mobile.

The dataset is hosted in a open repository on Github. This allows anyone to access and contribute by using the \emph{pull request} feature. We decided for this approach to allow the contribution to be reviewed before being accepted.
The contributor can also add to the catalog of problems and other meta-data like game platform, game genre, and game mode, allowing the dataset to keep evolving overtime.

About the limitations of the paper, this dataset is related only to postmortems problems, that is, things that ``went wrong''. Our plan is to also implement good practices gathered from the ``what went right'' sections in the future.
Also, the problems described by the authors are, in general, abstracts, without technical details. 

\section{Conclusion}
\label{sec:conclusion}

We presented a database of video-game development problems gathered from 200 postmortems. The database contains 1,035 problems categorized in three 20 different types.
It fills the gap between video-game development and software developers by providing a larger and more trustful information about video-game development problems. It thus provides a starting point for new researchers and practitioners to understand and propose solutions to video-game development problems.

Future plans for this work include 
(1) keep updating the database as more postmortems appear; 
(2) include other types of sources as, for example, technical blog posts and conference presentations like Game Developer Conference -- GDC; and 
(3) expanding the database by using the ``What went right'' section to gather \emph{good practices} instead of problems; 

This dataset can be used for researchers and practitioners to make a comparison with the common problems in traditional software development to better understand the differences in both domains and check the possibility of implementing traditional software engineering practices to mitigate those issues.
Also, the quotes in the problems should be further refined to more precisely define what are the causes of the problems. For example, what are the main causes of ``technical'' problems? Is it related to the game's type? Is there any correlation between the problems? From there we can draw common solutions from these issues.
Finally, investigate if the technical evolution (new hardware and tools) influence the type of problems in game development.

\section*{Acknowledgments}
The authors have been partly supported by the NSERC Discovery Grant program and Canada Research Chairs program.

\clearpage
\bibliographystyle{ACM-Reference-Format}
\bibliography{main.bib}


\begin{thebibliography}{13}


\ifx \showCODEN    \undefined \def \showCODEN     #1{\unskip}     \fi
\ifx \showDOI      \undefined \def \showDOI       #1{#1}\fi
\ifx \showISBNx    \undefined \def \showISBNx     #1{\unskip}     \fi
\ifx \showISBNxiii \undefined \def \showISBNxiii  #1{\unskip}     \fi
\ifx \showISSN     \undefined \def \showISSN      #1{\unskip}     \fi
\ifx \showLCCN     \undefined \def \showLCCN      #1{\unskip}     \fi
\ifx \shownote     \undefined \def \shownote      #1{#1}          \fi
\ifx \showarticletitle \undefined \def \showarticletitle #1{#1}   \fi
\ifx \showURL      \undefined \def \showURL       {\relax}        \fi
\providecommand\bibfield[2]{#2}
\providecommand\bibinfo[2]{#2}
\providecommand\natexlab[1]{#1}
\providecommand\showeprint[2][]{arXiv:#2}

\bibitem[\protect\citeauthoryear{{Ara Shirinian}}{{Ara Shirinian}}{2011}]%
        {Shirinian2011}
\bibfield{author}{\bibinfo{person}{{Ara Shirinian}}.}
  \bibinfo{year}{2011}\natexlab{}.
\newblock \bibinfo{title}{Dissecting The Postmortem: Lessons Learned From Two
  Years Of Game Development Self-Reportage}.
\newblock
  \bibinfo{howpublished}{\url{https://www.gamasutra.com/view/feature/6309/dissecting_the_postmortem_lessons_.php?print=1}}.
\newblock
\newblock
\shownote{[Online; accessed 5-March-2020].}


\bibitem[\protect\citeauthoryear{Blow}{Blow}{2004}]%
        {Blow2004}
\bibfield{author}{\bibinfo{person}{Jonathan Blow}.}
  \bibinfo{year}{2004}\natexlab{}.
\newblock \showarticletitle{{Game Development Harder Than You Think}}.
\newblock \bibinfo{journal}{\emph{Queue}} \bibinfo{volume}{1},
  \bibinfo{number}{10} (\bibinfo{date}{Feb.} \bibinfo{year}{2004}),
  \bibinfo{pages}{28}.
\newblock
\showISBNx{9781590598726}
\showISSN{1542-7730}
\urldef\tempurl%
\url{https://doi.org/10.1145/971564.971590}
\showDOI{\tempurl}


\bibitem[\protect\citeauthoryear{Callele, Neufeld, and Schneider}{Callele
  et~al\mbox{.}}{2005}]%
        {Callele2005}
\bibfield{author}{\bibinfo{person}{D. Callele}, \bibinfo{person}{E. Neufeld},
  {and} \bibinfo{person}{K. Schneider}.} \bibinfo{year}{2005}\natexlab{}.
\newblock \showarticletitle{Requirements engineering and the creative process
  in the video game industry}. In
  \bibinfo{booktitle}{\emph{13\textsuperscript{th} {IEEE} International
  Conference on Requirements Engineering ({RE}05)}}.
  \bibinfo{publisher}{{IEEE}}, \bibinfo{pages}{240--250}.
\newblock
\showISBNx{0769524257}
\urldef\tempurl%
\url{https://doi.org/10.1109/re.2005.58}
\showDOI{\tempurl}


\bibitem[\protect\citeauthoryear{Kanode and Haddad}{Kanode and Haddad}{2009}]%
        {Kanode2009}
\bibfield{author}{\bibinfo{person}{Christopher~M. Kanode} {and}
  \bibinfo{person}{Hisham~M. Haddad}.} \bibinfo{year}{2009}\natexlab{}.
\newblock \showarticletitle{Software engineering challenges in game
  development}, In \bibinfo{booktitle}{2009 Sixth International Conference on
  Information Technology: New Generations}.
\newblock \bibinfo{journal}{\emph{ITNG 2009 - 6\textsuperscript{th}
  International Conference on Information Technology: New Generations}},
  \bibinfo{pages}{260--265}.
\newblock
\showISBNx{9780769535968}
\urldef\tempurl%
\url{https://doi.org/10.1109/itng.2009.74}
\showDOI{\tempurl}


\bibitem[\protect\citeauthoryear{Lewis and Whitehead}{Lewis and
  Whitehead}{2011}]%
        {Lewis2011}
\bibfield{author}{\bibinfo{person}{Chris Lewis} {and} \bibinfo{person}{Jim
  Whitehead}.} \bibinfo{year}{2011}\natexlab{}.
\newblock \showarticletitle{The whats and the whys of games and software
  engineering}. In \bibinfo{booktitle}{\emph{Proceeding of the
  1\textsuperscript{st} international workshop on Games and software
  engineering - {GAS} 11}}. \bibinfo{publisher}{{ACM} Press},
  \bibinfo{pages}{1--4}.
\newblock
\showISBNx{9781450305785}
\showISSN{0270-5257}
\urldef\tempurl%
\url{https://doi.org/10.1145/1984674.1984676}
\showDOI{\tempurl}


\bibitem[\protect\citeauthoryear{Lin, Bezemer, and Hassan}{Lin
  et~al\mbox{.}}{2017}]%
        {Lin2017:updates}
\bibfield{author}{\bibinfo{person}{Dayi Lin}, \bibinfo{person}{Cor~Paul
  Bezemer}, {and} \bibinfo{person}{Ahmed~E. Hassan}.}
  \bibinfo{year}{2017}\natexlab{}.
\newblock \showarticletitle{{Studying the urgent updates of popular games on
  the Steam platform}}.
\newblock \bibinfo{journal}{\emph{Empirical Software Engineering}}
  \bibinfo{volume}{22}, \bibinfo{number}{4} (\bibinfo{year}{2017}),
  \bibinfo{pages}{2095--2126}.
\newblock
\showISSN{1573-7616}
\urldef\tempurl%
\url{https://doi.org/10.1007/s10664-016-9480-2}
\showDOI{\tempurl}


\bibitem[\protect\citeauthoryear{{Newzoo}}{{Newzoo}}{2019}]%
        {Newzoo19}
\bibfield{author}{\bibinfo{person}{{Newzoo}}.} \bibinfo{year}{2019}\natexlab{}.
\newblock \bibinfo{title}{2019 Global Games Market Report}.
\newblock
  \bibinfo{howpublished}{\url{https://newzoo.com/insights/trend-reports/newzoo-global-games-market-report-2019-light-version/}}.
\newblock
\newblock
\shownote{[Online; accessed 1-October-2019].}


\bibitem[\protect\citeauthoryear{Petrillo, Pimenta, Trindade, and
  Dietrich}{Petrillo et~al\mbox{.}}{2009}]%
        {Petrillo2009}
\bibfield{author}{\bibinfo{person}{F{\'{a}}bio Petrillo},
  \bibinfo{person}{Marcelo Pimenta}, \bibinfo{person}{Francisco Trindade},
  {and} \bibinfo{person}{Carlos Dietrich}.} \bibinfo{year}{2009}\natexlab{}.
\newblock \showarticletitle{What went wrong? a survey of problems in game
  development}.
\newblock \bibinfo{journal}{\emph{Computers in Entertainment}}
  \bibinfo{volume}{7}, \bibinfo{number}{1} (\bibinfo{date}{Feb.}
  \bibinfo{year}{2009}), \bibinfo{pages}{1}.
\newblock
\urldef\tempurl%
\url{https://doi.org/10.1145/1486508.1486521}
\showDOI{\tempurl}


\bibitem[\protect\citeauthoryear{Politowski, Fontoura, Petrillo, and
  Gu{\'e}h{\'e}neuc}{Politowski et~al\mbox{.}}{2016}]%
        {Politowski2016}
\bibfield{author}{\bibinfo{person}{Cristiano Politowski},
  \bibinfo{person}{Lisandra Fontoura}, \bibinfo{person}{Fabio Petrillo}, {and}
  \bibinfo{person}{Yann-Ga\"{e}l Gu{\'e}h{\'e}neuc}.}
  \bibinfo{year}{2016}\natexlab{}.
\newblock \showarticletitle{Are the Old Days Gone?: A Survey on Actual Software
  Engineering Processes in Video Game Industry}. In
  \bibinfo{booktitle}{\emph{Proceedings of the 5th International Workshop on
  Games and Software Engineering}} \emph{(\bibinfo{series}{GAS '16})}.
  \bibinfo{publisher}{ACM}, \bibinfo{pages}{22--28}.
\newblock
\showISBNx{978-1-4503-4160-8}
\urldef\tempurl%
\url{https://doi.org/10.1145/2896958.2896960}
\showDOI{\tempurl}


\bibitem[\protect\citeauthoryear{Politowski, Fontoura, Petrillo, and
  Gu{\'{e}}h{\'{e}}neuc}{Politowski et~al\mbox{.}}{2018}]%
        {Politowski2018}
\bibfield{author}{\bibinfo{person}{Cristiano Politowski},
  \bibinfo{person}{Lisandra~M. Fontoura}, \bibinfo{person}{Fabio Petrillo},
  {and} \bibinfo{person}{Yann~Ga{\"{e}}l Gu{\'{e}}h{\'{e}}neuc}.}
  \bibinfo{year}{2018}\natexlab{}.
\newblock \showarticletitle{{Learning from the past: A process recommendation
  system for video game projects using postmortems experiences}}.
\newblock \bibinfo{journal}{\emph{Information and Software Technology}}
  \bibinfo{volume}{100}, \bibinfo{number}{March} (\bibinfo{year}{2018}),
  \bibinfo{pages}{103--118}.
\newblock
\showISSN{0950-5849}
\urldef\tempurl%
\url{https://doi.org/10.1016/j.infsof.2018.04.003}
\showDOI{\tempurl}


\bibitem[\protect\citeauthoryear{{Stol}, {Ralph}, and {Fitzgerald}}{{Stol}
  et~al\mbox{.}}{2016}]%
        {Stol2016}
\bibfield{author}{\bibinfo{person}{K. {Stol}}, \bibinfo{person}{P. {Ralph}},
  {and} \bibinfo{person}{B. {Fitzgerald}}.} \bibinfo{year}{2016}\natexlab{}.
\newblock \showarticletitle{Grounded Theory in Software Engineering Research: A
  Critical Review and Guidelines}. In \bibinfo{booktitle}{\emph{2016 IEEE/ACM
  38th International Conference on Software Engineering (ICSE)}}.
  \bibinfo{pages}{120--131}.
\newblock
\showISSN{1558-1225}
\urldef\tempurl%
\url{https://doi.org/10.1145/2884781.2884833}
\showDOI{\tempurl}


\bibitem[\protect\citeauthoryear{Tschang}{Tschang}{2005}]%
        {Tschang2005}
\bibfield{author}{\bibinfo{person}{F~Ted Tschang}.}
  \bibinfo{year}{2005}\natexlab{}.
\newblock \showarticletitle{Videogames as interactive experiential products and
  their manner of development}.
\newblock \bibinfo{journal}{\emph{International Journal of Innovation
  Management}} \bibinfo{volume}{9}, \bibinfo{number}{01}
  (\bibinfo{year}{2005}), \bibinfo{pages}{103--131}.
\newblock


\bibitem[\protect\citeauthoryear{Washburn, Sathiyanarayanan, Nagappan,
  Zimmermann, and Bird}{Washburn et~al\mbox{.}}{2016}]%
        {Washburn2016}
\bibfield{author}{\bibinfo{person}{Michael Washburn}, \bibinfo{person}{Pavithra
  Sathiyanarayanan}, \bibinfo{person}{Meiyappan Nagappan},
  \bibinfo{person}{Thomas Zimmermann}, {and} \bibinfo{person}{Christian Bird}.}
  \bibinfo{year}{2016}\natexlab{}.
\newblock \showarticletitle{What Went Right and What Went Wrong: An Analysis of
  155 Postmortems from Game Development}, In \bibinfo{booktitle}{Proceedings of
  the 38\textsuperscript{th} International Conference on Software Engineering
  Companion - {ICSE} 16}.
\newblock \bibinfo{journal}{\emph{Proceedings of the 38\textsuperscript{th}
  International Conference on Software Engineering (ICSE 2016 SEIP Track)}}.
\newblock
\showISBNx{9781450342056}
\urldef\tempurl%
\url{https://doi.org/10.1145/2889160.2889253}
\showDOI{\tempurl}


\end{thebibliography}
\end{document}